\documentclass[usenatbib, useAMS]{mn2e}
\usepackage{times}
\usepackage{color}
\usepackage{tensor}
\usepackage{multirow}
\usepackage{graphicx}
\usepackage{amsmath}
\usepackage{natbib}
\usepackage{myaasmacros}
\usepackage[english]{babel}
\usepackage{mathptmx}
\usepackage{tabularx}
\usepackage{placeins}

\newcommand{\Msol}{\mathrm{M}_{\odot}}
\newcommand{\kpc}{\mathrm{kpc}}
\newcommand{\Mpc}{\mathrm{Mpc}}
\newcommand{\kks}{\kpc\,\mathrm{km\,s^{-1}}}
\newcommand{\ks}{\mathrm{km^2\,s^{-2}}}
\newcommand{\Phie}{\Phi_{\mathrm{eff},a}}
\def\lsim{\mathrel{\rlap{\lower3pt\hbox{$\sim$}}
    \raise1pt\hbox{$<$}}}                % less than or approx. symbol
\def\gsim{\mathrel{\rlap{\lower3pt\hbox{$\sim$}}
    \raise1pt\hbox{$>$}}}                % greater than or approx. symbol

\begin{document}
\pubyear{2012}
\title[Maximum entropy dark matter halos]{Conserved actions, maximum entropy and dark
  matter halos}
\author[A. Pontzen \& F. Governato]{Andrew
  Pontzen$^{1,2,3}$, Fabio Governato$^{4}$ \\ 
$^1$ {Oxford Astrophysics, Denys Wilkinson Building, Keble Road,
  Oxford, OX1 3RH} \\
$^2$ {Balliol College, Broad Street, Oxford, OX1 3BJ} \\
$^3$ {Email: andrew.pontzen@astro.ox.ac.uk} \\
$^4$ {Astronomy Department, University of Washington, Seattle, WA 98195, USA}}
\date{ Received ---; published---. }
\maketitle

\newcommand{\pc}{\mathrm{pc}}
\newcommand{\cm}{\mathrm{cm}}
\newcommand{\Myr}{\mathrm{Myr}}
\newcommand{\dd}{\mathrm{d}}
\newcommand{\Gyr}{\mathrm{Gyr}}
\newcommand{\ergs}{\mathrm{ergs}}
\vspace{-1cm}
\begin{abstract}
  We use maximum entropy arguments to derive the phase space
  distribution of a virialized dark matter halo. Our distribution
  function gives an improved representation of the end product of
  violent relaxation. This is achieved by incorporating physically
  motivated dynamical constraints (specifically on orbital
  actions) which prevent arbitrary redistribution of energy.

We compare the predictions with three high-resolution dark matter
simulations of widely varying mass. The numerical distribution
function is accurately predicted by our argument, producing an
excellent match for the vast majority of particles.

The remaining particles constitute the central cusp of the halo
($\lsim 4\%$ of the dark matter). They can be accounted for within the
presented framework once the short dynamical timescales of the
centre are taken into account.

\vspace{0.2cm}
\end{abstract}

\section{Introduction}

For over two decades it has been possible to use numerical methods to
model systems of cold, collisionless dark matter particles collapsing
under gravity to form stable, virialized `halos'
(\citeauthor{1985Natur.317..595F} 1985;
\citeauthor{1991ApJ...378..496D} 1991; for a review see
\citeauthor{2012arXiv1210.0544F} 2012). The density of these
halos declines with radius following a slowly changing power law
dependence, roughly $\rho \sim r^{-1}$ at small radii and
$\rho \sim r^{-3}$ in the outer regions
\citep{1996ApJ...462..563N,1997ApJS..111...73K,1998ApJ...499L...5M}.
Despite some early uncertainty, recent simulations with independent
computer codes all reproduce this result
\cite[e.g.][]{2008Natur.454..735D,2010MNRAS.402...21N,2009MNRAS.398L..21S}.
The universal behaviour seems to be independent of the power spectrum
of initial linear density fluctuations
\citep{1999MNRAS.310.1147M,2005MNRAS.357...82R,2009MNRAS.396..709W} as
well as the mass of the collapsed object and the epoch of collapse,
although together these determine a scale radius for the transition
from $r^{-1}$ to $r^{-3}$ behaviour
\citep{1996MNRAS.281..716C,1997ApJ...490..493N,2001MNRAS.321..559B,2001ApJ...554..114E,2007MNRAS.378...55M}.
Even simulations of `cold collapse', for which initial conditions
consist of a homogeneous sphere of particles, seem to produce similar
universal profiles \citep{1999ApJ...517...64H}.

It remains an outstanding question, however, whether this universality
can be adequately explained from first principles. Until this question
is answered, we do not fully understand what the universality means
and must rely on new simulations to predict the effect of changes in
the initial conditions or particle properties. The experimental result
that monolithic collapse produces the same types of
system as hierarchical merging
\citep{1999ApJ...517...64H,1999MNRAS.310.1147M,2009MNRAS.396..709W} is
provocative: it means that any explanation for universality which
invokes a specific cosmology
\citep[e.g.][]{1998MNRAS.293..337S,2003ApJ...588..680D,2012MNRAS.423.2190S}
must be describing a special case of a more general process
\citep{2003ApJ...593...26M}.

Attempts to understand collisionless gravitational collapse's
insensitivity to initial conditions were pioneered by
\cite{1967MNRAS.136..101L} in the context of self-gravitating stellar
systems. Adopting Boltzmann's procedure for deriving the
thermodynamics of collisional systems, Lynden-Bell maximized the
entropy of the systems subject to fixed energy. This implies a density
profile obeying $\rho(r) \propto r^{-2}$, so disagrees with the results
of numerical experiments.  Moreover this approach gives rise to a
number of physically questionable conclusions \citep[for reviews
see][]{1990PhR...188..285P,1999PhyA..263..293L}. The clearest of these
is the `gravothermal catastrophe': entropy can be increased without
bound by transferring energy from the innermost orbits of a
self-gravitating system to the outermost orbits
\citep{1968MNRAS.138..495L,1986MNRAS.219..285T}. This implies the
existence of a runaway physical instability in which the majority of material
collapses into an extreme central density cusp or black hole. 

Observations and numerics both suggest that something prevents the
above catastrophe from occurring on any reasonable timescale; in other
words, a physical constraint is preventing the arbitrary
redistribution of energy \citep{1987MNRAS.229..103W}. In the spirit of
\cite{1957PhRv..106..620J} a general explanation for the final state
can still be based on the ideas of statistical mechanics but in the
presence of constraints other than energy. The additional constraints
will represent the incompleteness of the energy equilibriation effects
of violent relaxation.

This is the approach we adopt in the present work. The likely
distribution of particles in phase space is selected by maximizing
entropy,
\begin{equation}
  S= -k \int f\, \ln f\, \dd^6 \omega\textrm{,}\label{eq:entropy}
\end{equation}
where $k$ is Boltzmann's constant and $f(\omega)$ is the probability
of finding a particle in a specified region of phase space $\omega$,
subject to relevant constraints on the desired solution. (This
approach can be motivated by showing that the vast majority of states
consistent with a given set of constraints are to be found near the
maximum entropy solution; see Appendix \ref{sec:but-what-about} for
further discussion and references.)

The constraints applied arise from the dynamical evolution of
collisionless systems. We will argue that, in the late stages of
violent relaxation, there is a diffusion of particles in phase space
which approximately conserves the sum of orbital actions. This sum is
progressively better conserved as equilibrium is approached and
therefore its role in establishing that equilibrium cannot be
ignored. Applying the maximum entropy recipe, we will show that the
phase space structure of equilibrium halos is then reproduced over
orders of magnitude in probability density. To our knowledge, this is
the first instance of maximum entropy reasoning, applied to 6D phase
space and subject to physically motivated constraints, producing such
success in a collisionless system.

The remainder of this work is organized as follows. First, the
conservation of action is discussed (Section
\ref{sec:conservation-action}). Then a canonical ensemble constructed
on this basis (Section \ref{sec:new-canon-ensemble}) yields a phase
space distribution in quantitative agreement with high resolution
numerical experiments (Section \ref{sec:simulations}). A discrepancy
affecting a small fraction of particles at low angular momentum is
highlighted in Section \ref{sec:results-j}. Finally we discuss the
predicted radial density profiles which again highlight the need for
special treatment of low angular momentum orbits (Section
\ref{sec:real-space}). We conclude in Section \ref{sec:conclusions}.

\section{The analytic phase space distribution}

\subsection{Conservation of action}\label{sec:conservation-action}

Our first task is to identify and explain some relevant quantities
which should be held fixed when maximizing entropy.  This will be
central to our argument because such constraints represent the
incompleteness of violent relaxation's tendency to redistribute
energy, generating a different solution from the one based on energy
conservation alone.  With this in mind we will show that the radial
action $J_r$ (to be defined below) is conserved in an average sense
even during rapid potential changes.

This average conservation does not appear to have been discussed
elsewhere in the literature. We will first show how it
can be derived from previous work \citep{2012MNRAS.421.3464P} when the
potential changes instantaneously, maintaining the sphericity of the
halo. Then a more general (but more abstract) argument will be given
which additionally shows that the other two actions (the $z$-component
of the angular momentum $j_z$ and the scalar angular momentum $j$) are
also conserved in the same average sense. The second approach
encompasses perturbations to the potential which have variations on
arbitrary timescales and may break spherical symmetry. However the
first has a more intuitive content and therefore forms our starting
point.

In a spherical system, the radial action $J_r$ is defined by
\begin{equation}
J_r = \frac{1}{\pi} \int_{r_{\mathrm{min}}}^{r_{\mathrm{max}}} \sqrt{2E-2\Phi(r; t) - j^2/r^2}\, \dd r\textrm{,}\label{eq:radial-action}
\end{equation}
Here $E$ is the specific energy, $j$ is the specific angular momentum and
$\Phi$ is the potential at a given radius $r$ and time $t$; the $r$
integral is taken over the region where the integrand is real. 

The radial action $J_r$ has the same units as specific angular
momentum $j$. This reflects the similar conservation roles these two
quantities play for the radial and angular components of the
motion. In particular $J_r$ is exactly conserved if any changes in the
potential occur sufficiently slowly (`adiabatically') in time
\citep[e.g.][]{1987gady.book.....B}.

On the other hand in the rapid, impulsive limit under an instantaneous
change in energy $E \to E + \Delta E$ and potential $\Phi(r) \to
\Phi(r) + \Delta \Phi(r)$, the action of the particle is changed at
first order:
\begin{equation}
\Delta J_r = \left.\frac{\partial J_r}{\partial
    E}\right|_{\Phi} \Delta E + \int_0^{\infty} \dd r\, \Delta \Phi
\left. \frac{\delta J_r}{\delta \Phi} \right|_{E}\textrm{.} \label{eq:delta-J-delta-E-relation}
\end{equation}
In \cite{2012MNRAS.421.3464P} we showed that the energy shift
$\Delta E$ induced by the change of potential, averaged over possible
orbital phases of the particle, is
\begin{equation}
\langle \Delta E \rangle = - \frac{\int_0^{\infty} \dd r \,\Delta \Phi \left. \delta J_r / \delta \Phi
  \right|_{E}}{\left. \partial J_r
  / \partial E \right|_{\Phi}}\textrm{,}\label{eq:average-energy-shift}
\end{equation}
an exact result 
\cite[see equation 12 of][]{2012MNRAS.421.3464P}. Here angular
brackets denote averaging over all possible phases of the orbit. Considering the probability
distribution of radial actions after this change, one has $\langle
\Delta J_r \rangle = 0$ at first order, by substituting equation
\eqref{eq:average-energy-shift} in
\eqref{eq:delta-J-delta-E-relation}. Even though a specific particle
will change its radial action, the ensemble average is conserved.

This result connects closely with the standard adiabatic argument that
$\Delta J_r=0$ if any changes to the potential occur on long
timescales. In our case, however, the necessary `phase averaging' does
not occur over time for an individual particle but instead via a statistical
consideration of an ensemble of particles spread evenly through all
possible phases.

\begin{figure}
\includegraphics[width=0.49\textwidth]{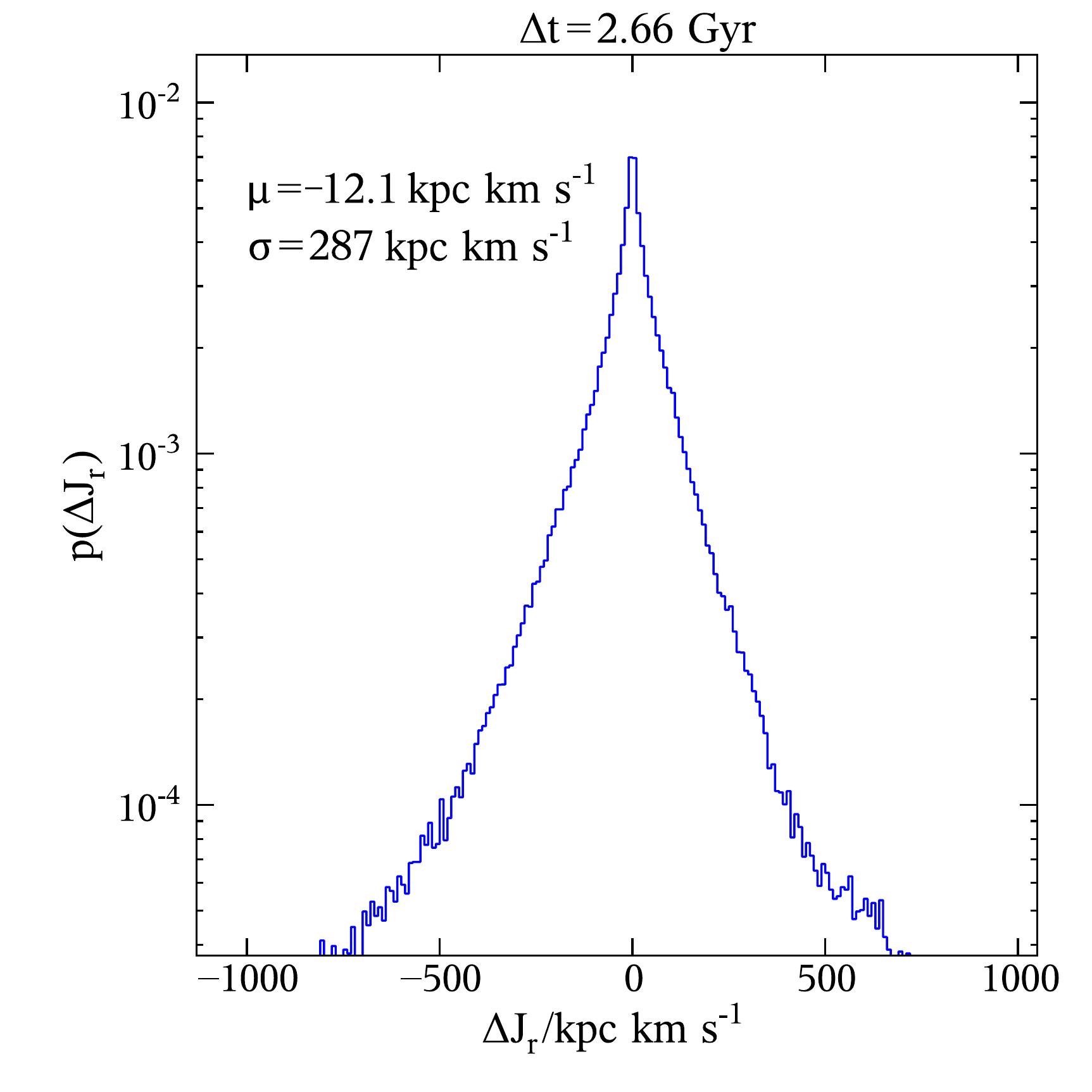}
\caption{An illustration of the diffusion of particles in action
  space. Here particles in the inner $10\,\kpc$ of a simulation have been
  selected and their radial actions $J_r$ numerically calculated at
  two timesteps separated by $\Delta t=2.7\,\mathrm{Gyr}$. The change
  in the population mean action is small ($\mu = -12.1\kks$) compared
  against the magnitude of the random diffusion ($\sigma
  =287\,\kks$). }\label{fig:diffuse}
\end{figure}

We can generalize as follows.  Adopting the complete set of
action-angle coordinates for phase space \citep[e.g.][]{1979LNP...110.....P,1987gady.book.....B}, the momenta are
$\vec{J}=(J_r, j, j_z)$ where $j$ is the total angular momentum and
$j_z$ is its component in the $z$ direction (so $-j<j_z<j$). The
conjugate coordinates are $\vec{\Theta}=(\psi_r, \phi, \chi)$, taken
to be periodic with interval $2 \pi$, and the Hamiltonian is
\begin{equation}
H(\vec{J}, \vec{\Theta}) = H_0(\vec{J}, t) +  h(\vec{J}, \vec{\Theta}, t)\textrm{.}
\end{equation}
Here $h$ is an arbitrary perturbation. It may consist of a long-lived
term (perhaps a departure from spherical symmetry) and fluctuations on
arbitrary  time-scales. In the background, the $\vec{\Theta}$
coordinates change at a constant rate,
\begin{equation}
\vec{\Theta} = \vec{\Theta}_0 + \vec{\Omega} \, t\textrm{,}\label{eq:theta-of-t}
\end{equation}
and the frequencies for the radial and azimuthal motion $\Omega_r$ and
$\Omega_j$ obey
\begin{equation}
\Omega_r = \frac{\partial{H_0}}{\partial J_r}; \hspace{0.5cm} \Omega_j
= \frac{\partial H_0}{\partial j},\label{eq:omegas}
\end{equation}
by Hamilton's equations. The frequencies may change slowly with time
($\dd \Omega /\dd t \ll \Omega^2$). $\vec{J}$ is conserved
in the background but when $h$ is non-zero,
the equations of motion read
\begin{equation}
\frac{\dd \vec{J}}{\dd t} = - \frac{\partial{h}}{\partial\vec{\Theta}}\textrm{.}
\end{equation}
Because equation \eqref{eq:theta-of-t} shows that particles in the
background move at a uniform rate in the $\vec{\Theta}$ coordinates, an
equilibrium distribution $f$ has no $\vec{\Theta}$
dependence; {\it i.e.} the density $f$ is a function of $\vec{J}$
alone\footnote{The conjugate coordinate to $j_z$ is also a constant of
  motion in the background, so this argument does not strictly show
  $\dd \langle j_z  \rangle/\dd t =0$.
  However $\langle j_z \rangle$ must be exactly conserved anyway if
  the perturbations are internally generated, since it is proportional
  to a
  component of the total angular momentum vector.}. From this it follows that
\begin{equation}
\frac{\dd}{\dd t} \langle \vec{J} \rangle = -\int \dd^3 J \dd^3 \Theta
f(\vec{J}) \frac{\partial h}{\partial \vec{\Theta}} = 0\textrm{,}
\end{equation}
where the result is obtained via integration by parts.  This means an
individual particle's $\vec{J}$ can `diffuse'
\citep{1988MNRAS.230..597B} over large distances $\sigma$ in action space,
\begin{equation}
\sigma \equiv \langle (\delta J)^2 \rangle^{1/2} = \mathcal{O}(\epsilon)\textrm{,}
\end{equation}
compared to the variation in the mean $\mu$ of the population,
\begin{equation}
\mu \equiv \langle \delta J \rangle = \mathcal{O}(\epsilon^2).
\end{equation}
We can inspect this diffusion in a simulation by calculating the
relevant actions at two timesteps. As an example, Figure
\ref{fig:diffuse} shows a histogram of changes in the radial action of
tightly bound particles in the forming ``Dwarf'' halo (see Section
\ref{sec:simulations}) between $z=3.1$ and $z=1.4$, a time interval of
approximately $2.7\,\Gyr$. We define `tightly bound' by selecting
particles interior to $10\,\kpc$ at the earlier time step, and
calculate the $J_r$ values of these particles in both outputs
according to the numerical recipes given later (Section
\ref{sec:perf-comp-techn}).

As expected from the linear analysis, Figure \ref{fig:diffuse} shows
that individual simulated particles change their actions more rapidly
than the population mean. Quantitatively, the change in the population
mean $\mu$ is $-12.1\,\kks$ whereas a typical particle has moved by
$\sigma=287\,\kks\gg |\mu|$ from its original $J_r$ value. Note also
that the mean $J_r$ value for these particles in the final timestep is
$\langle J_r \rangle = 187\,\kks<\sigma$, meaning particles really do
cover significant distances in action space.

This analysis confirms that $\langle \vec{J} \rangle$ evolves slowly
and, although it is not exactly conserved, it forms a constraint of
motion that cannot be ignored on finite timescales. A complete
description would require investigation of different moments of the
distribution at higher order in perturbation theory. For now, however,
we have motivated a picture in which $\langle \vec{J} \rangle$ evolves
sufficiently slowly compared to the diffusion of an individual
particle that it must be considered fixed in analysis of the
distribution.

\subsection{The new canonical ensemble}\label{sec:new-canon-ensemble}

In the Introduction we explained that, to obtain a phase space
distribution function, we will maximize the entropy~\eqref{eq:entropy}
subject to constraints on the particle population (further discussion
is given in Appendix \ref{sec:but-what-about}).  As well as energy
conservation, we apply the 3-vector of constraints on $\langle \vec{J}
\rangle$ discussed above. This gives rise to a total of four Lagrange
multipliers in the resulting distribution function:
\begin{equation}
f(\vec{J}) \propto \exp \left(-\vec{\beta} \cdot \vec{J} -\beta_E E(\vec{J})\right)\textrm{,}\label{eq:maxent-solution}
\end{equation}
where the Lagrange multipliers are $\vec{\beta} = (\beta_r, \beta_j,
\beta_z)$ and $\beta_E$. In the absence of the new constraints,
$\vec{\beta}=0$ and $\beta_E$ is identified with $1/kT$ (where $T$ is
the thermodynamical temperature).

All four constants can be determined in a
variety of ways depending on the situation; for a complete account of
structure formation one would like to be able to derive them from the
initial conditions, but this lies beyond the scope of the current
paper (although see Section \ref{sec:other-simulations} for further comments).
The lack of any reference to $\vec{\Theta}$ in
equation~\eqref{eq:maxent-solution} indicates that the solution is
phase-mixed, as required for equilibrium.

Equation~\eqref{eq:maxent-solution} is the essential prediction of the
present work. As with any prediction derived from a maximum entropy
argument, it will be able to fit the actual ensemble only if we have
encapsulated enough of the dynamics within the constraints
\citep{Jaynes79}. The rest of this paper is concerned with testing to
what extent that is the case.

\begin{figure*}
\begin{center}
\includegraphics[width=18cm]{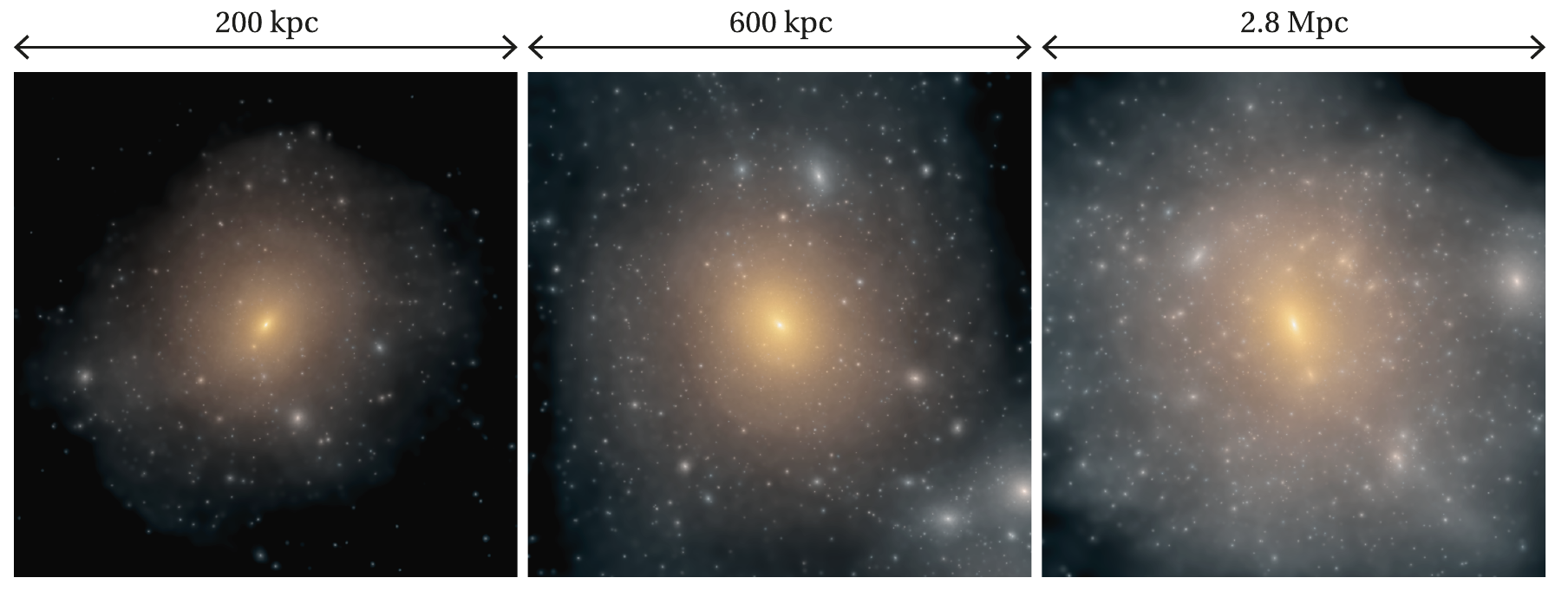}
\begin{tabularx}{17.8cm}{XXX}
\textbf{Dwarf} &  \textbf{MW} & \textbf{Cluster}  \\
$r_{200} = 98\,\kpc$; $M_{200}=2.8 \times 10^{10}\,\Msol$; $c=19.6$ &
$r_{200} = 301\,\kpc$; $M_{200}=8.0 \times 10^{11}\,\Msol$; $c=15.5$ &
$r_{200} = 1.43\,\Mpc$;  $M_{200} = 8.7\times10^{13}\,\Msol$; $c=9.9$ \\ 
$N_{\mathrm{part}} = 3.4 \times 10^6$; $\epsilon = 65\,\pc$. &
$N_{\mathrm{part}} = 5.3 \times 10^6$; $\epsilon = 170\,\pc$. &
$N_{\mathrm{part}} = 8.9 \times 10^6$; $\epsilon = 690\,\pc$.
\end{tabularx}
\end{center}
\caption{Images of the three dark matter simulations, accompanied
  by numerical properties. Respectively $r_{200}$, $M_{200}$, $c$,
  $N_{\mathrm{part}}$ and $\epsilon$ denote the radius at which the
  density exceeds the critical density by a factor $200$; the mass
  within this radius; the `concentration', $c=r_{200}/r_s$ where $r_s$
  is the NFW scale radius as described in the text; the number of
  particles within $r_{200}$; and the gravitational softening length
  in physical units at $z=0$.  The images are scaled to show the
  virial sphere of the main halo. The brightness represents the column
  density of dark matter (scaled logarithmically to give a dynamic
  range of $3000$ in each case); the colour corresponds to a
  density-weighted potential along the line of
  sight. }\label{fig:simulations}
\end{figure*}
First consider the probability of finding a particle with $J_r$ in a
given interval (ignoring $j$ and $j_z$ coordinates). This is given by
\begin{equation}
p_r(J_r) = \int_0^{2\pi} \dd^3 \Theta \int_0^{\infty} \dd j \int_{-j}^j
\dd j_z\, f(\vec{J})\textrm{,}\label{eq:marginalize-out-j}
\end{equation}
because the action-angle coordinates are canonical, so the phase-space
measure is constant. In the limit that the energy of the system
becomes large at fixed action, equation \eqref{eq:marginalize-out-j} can be solved:
\begin{equation}
p_r(J_r) \propto \exp - \beta_r J_r\hspace{1cm}(\beta_E=0)\textrm{,}\label{eq:Jr-only-distrib}
\end{equation}
but it is not immediately clear whether we will be operating in
this regime. More generally a closed form for $p_r(J_r)$ is hard to
obtain, but we can at least show that
\begin{equation}
\frac{\dd \ln p_r}{\dd J_r} = -\beta_E \langle \Omega_r
\rangle_{J_r} - \beta_r\textrm{,}\label{eq:MAXENT-Jr-derivative}
\end{equation}
which can then be integrated numerically for a given case to give a
concrete comparison between equation \eqref{eq:maxent-solution} and
simulations. Here we have defined
\begin{equation}
\langle \Omega_r \rangle_{J_r} \equiv \frac{1}{p_r(J_r)}\int f(\vec{J})
\Omega_r(\vec{J})\, \dd j\, \dd j_z\textrm{,}
\end{equation}
which is the mean of the radial frequency for particles with a fixed
$J_r$.  With this definition, relation \eqref{eq:MAXENT-Jr-derivative} can be derived
from equation \eqref{eq:marginalize-out-j}, recalling that the radial
frequency $\Omega_r$ of the particle's orbit obeys
equation~\eqref{eq:omegas}.  We will investigate and explain the
distribution of $J_r$ values predicted by
equation~\eqref{eq:MAXENT-Jr-derivative} in Section \ref{sec:results}.

Now consider the distribution of total angular momentum $j$. We will
follow exactly the same series of manipulations as for the radial
action; however in this case the marginalization over $j_z$ introduces
a non-trivial term:
\begin{align}
p_j(j) & = \int_0^{2 \pi} \dd^3 \Theta \int_{-j}^{j} \dd j_z\, \int_0^{\infty} \dd J_r\,
f(\vec{J}) & \nonumber \\ 
& \propto \sinh \left( \beta_z j \right) \, \exp \left(-\beta_j j \right) \int_0^{\infty} \dd J_r
\exp \left( - \beta_E E \right) \textrm{.}\label{eq:pj-function}
\end{align}
Once again, in the case $\beta_E\to 0$, we have a fully analytic
expression for $p_j(j)$\textrm{,}
\begin{equation}
p_j(j) \propto \sinh \left(\beta_z j \right) \exp \left( - \beta_j j \right)\hspace{1cm}(\beta_E=0)\textrm{,}\label{eq:angmom-no-E}
\end{equation}
which will serve as a useful point of comparison. More generally we
can differentiate equation \eqref{eq:pj-function} to obtain
\begin{equation}
\frac{\dd \ln p_j}{\dd j} = \beta_z \coth \left(\beta_z j\right) -\beta_E \langle \Omega_j \rangle_j - \beta_j\textrm{,}\label{eq:MAXENT-angmom}
\end{equation}
where $\Omega_j$ is the angular frequency of the orbit and
\begin{equation}
\langle \Omega_j \rangle_j \equiv \frac{1}{p_j(j)} \int f(\vec{J})
\Omega_j(\vec{J}) \dd J_r \, \dd j_z
\end{equation}
is the mean angular frequency of particles at fixed $j$. Equation
\eqref{eq:MAXENT-angmom} for the angular momentum distribution
(ignoring all other coordinates) is the equivalent of equation
\eqref{eq:MAXENT-Jr-derivative} for the radial action
distribution. Once again we will investigate and explain the shape it
predicts in Section~\ref{sec:results-j}.  First, however, we will
explain the simulations which serve as a point of comparison for the
later discussions.

\section{Comparison to simulations}\label{sec:simulations}

\subsection{Overview of the simulations}

\begin{figure*}
\includegraphics[width=0.9\textwidth]{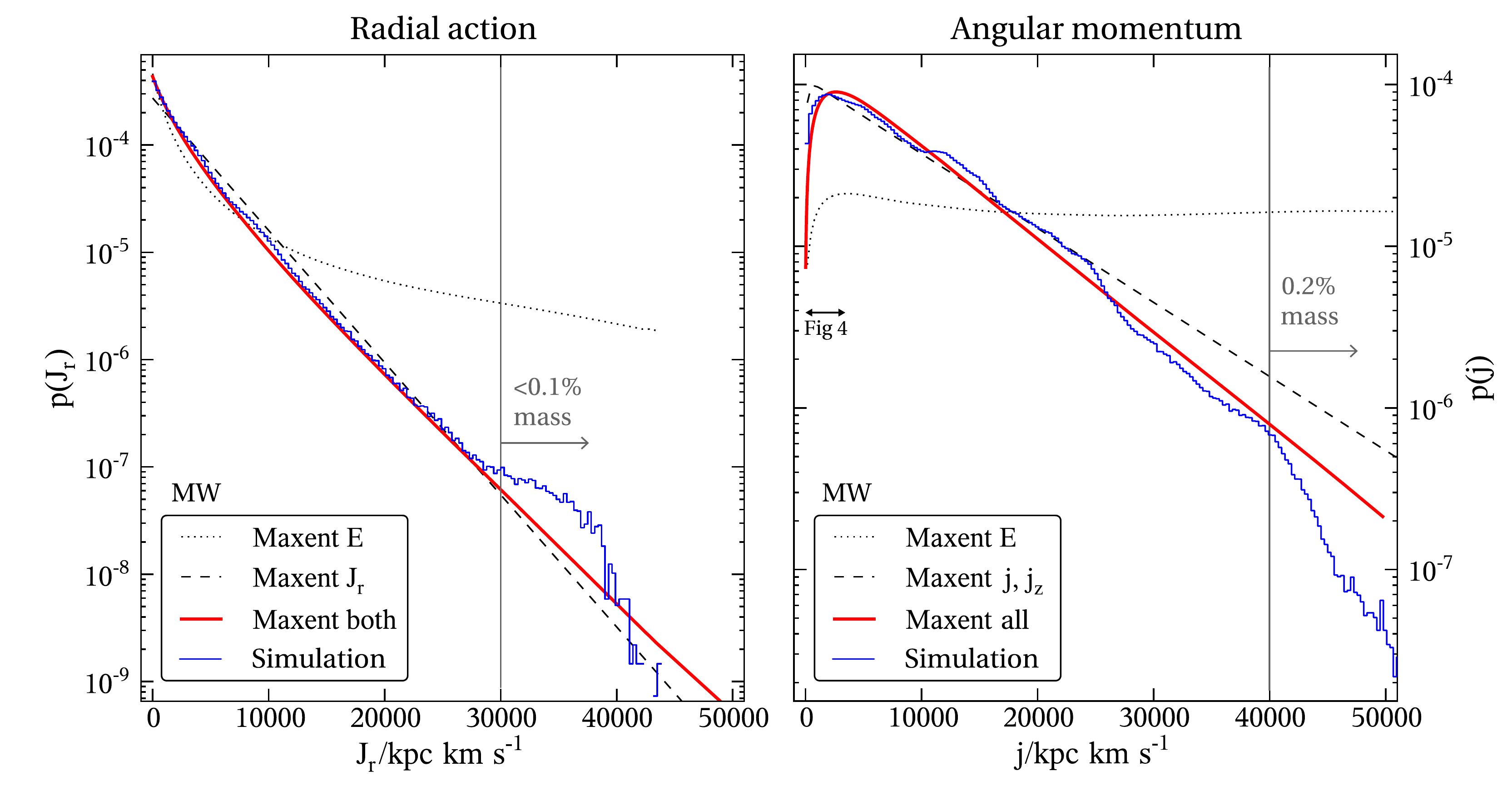}
\caption{The distribution of particles' radial action $J_r$ (left
  panel) and scalar angular momentum $j$ (right panel) in MW (both
  panels show the distribution of particles as a
  histogram). Also shown are maximum entropy solutions based on energy
  conservation alone (dotted curve); action conservation alone (dashed
  curve) and our advocated solution using both constraints (thick solid curve). The last of these
  provides a good reproduction of the distribution for
  $J_r<J_{r,\mathrm{break}} = 3 \times 10^4\,\mathrm{kpc\,km\,s^{-1}}$
  and $j<j_{\mathrm{break}}=4\times 10^4\,\mathrm{kpc\,km\,s^{-1}}$
  respectively, while extending over orders of magnitude in
  probability density. Less than $0.1\%$ of particles lie at
  $J_r>J_{r,\mathrm{break}}$; approximately $0.2\%$ lie at
  $j>j_{\mathrm{break}}$. Despite the good overall agreement, problems
  become apparent at very small $j$; a blowup of the indicated range
  $j<3500\, \kks$ is given in Figure 4. }\label{fig:Jr}
\end{figure*}

In the previous section we applied maximum entropy reasoning to
conservation of energy and approximate conservation of action
to derive an expected equilibrium phase space distribution. We
will now compare that expectation against simulated dark
matter halos. Our strategy is to integrate equations
\eqref{eq:MAXENT-Jr-derivative} and \eqref{eq:MAXENT-angmom}
numerically for these simulations and compare to the actual
distribution of particles binned by $J_r$ and $j$ respectively.

We will present results from three simulated dark matter halos (shown
in Figure~\ref{fig:simulations}), chosen to span a wide range of
masses with an approximately constant number of particles per halo
(several million in each case). We also compared our results against
the GHALO multi-billion-particle phase space
\citep{2009MNRAS.398L..21S}, finding good agreement similar to that
described for our ``MW'' halo here. This gives confidence that the
mechanisms and results discussed in the paper are not sensitive to
numerical resolution.

Our simulations are run from cosmological initial conditions at
$z\simeq 100$ in a `zoom' configuration \citep{1993MNRAS.265..271N},
i.e. with high resolution for the main halo and its immediate
surroundings and lower resolution for the cosmological
environment. The softening lengths $\epsilon$ for the high resolution
region are listed in Figure~\ref{fig:simulations} and are fixed in
physical units from $z=9$, prior to which they scale linearly with
cosmological scalefactor, a compromise motivated by numerical
convergence studies \citep{2004MNRAS.348..977D}. We verified at the
final output ($z=0$) that the high resolution regions have not been
contaminated by low resolution particles, and that the halo real-space
density profiles are well described by a slowly rolling powerlaw, in
accordance with all recent simulations \citep[e.g.][and references
therein]{2008Natur.454..735D,2009MNRAS.398L..21S,2010MNRAS.402...21N}.

Each
simulation output contains full cartesian phase space coordinates
($\vec{x}$, $\vec{v}$). The position space is re-centred on the
central density peak of the halo using the `shrinking sphere' method
of \cite{2003MNRAS.338...14P}. The velocities are re-centred such that
a central sphere of radius $r_{200}/30$ has zero net velocity, where 
$r_{200}$ is the radius at which the mean halo density is $200$ times
the critical density. Henceforth we only consider particles inside $r_{200}$.

From left to right in Figure \ref{fig:simulations} the simulated halos
become more massive. The width of each panel is equal to $2r_{200}$
and the luminosity is scaled to represent the column density over a
dynamic range of $3000$. The most
conspicuous aspect of Figure 2 is that the halos become less centrally
concentrated. We verified this by fitting a classic ``NFW''
\citep{1996ApJ...462..563N} formula to the density profile. The NFW
fit,
\begin{equation}
\rho(r) = \frac{\rho_0}{(1+r/r_s)^2\,(r/r_s)}\textrm{,}
\end{equation}
yields $\rho_0$, a characteristic density, and $r_s$, a scale
radius. The latter is often expressed in a scale-free manner as a
concentration value $c=r_{200}/r_s$; we have recorded the value for
each halo in Figure \ref{fig:simulations}. As expected the
concentration decreases with increasing mass, in agreement with
previously known trends
\citep[e.g.][]{2007MNRAS.378...55M,2001MNRAS.321..559B}. We thus have
a sample of cosmological halos which span a wide range in both mass
and concentration.  These different concentrations are thought to arise from
different mean densities in the universe at the epoch of collapse
\citep{1997ApJ...490..493N,2001MNRAS.321..559B}.
\label{sec:perf-comp-techn}

All halos in Figure \ref{fig:simulations} exhibits large amounts of
substructure; we will present results with this substructure
subtracted, although we have verified that including the
substructure does not have a qualitative impact on our results. The
substructure is identified and removed using the ``Amiga Halo Finder''
\citep{2009ApJS..182..608K}.  For each remaining particle inside
$r_{200}$, the specific scalar angular momentum is given by
$j=\left|\vec{v} \times \vec{x}\right|$. We calculate $J_r$ by
evaluating equation~\eqref{eq:radial-action} numerically, using a
spherically-averaged potential $\Phi$ defined by
\begin{equation}
\Phi(r) = \int_0^r \dd r' \frac{GM(<r')}{r'^2}\textrm{,}\label{eq:Phi-equation}
\end{equation}
where $M(<r')$ is the total mass enclosed by a sphere of radius $r'$,
and the specific energy $E$ of each particle is defined as
$E=\vec{v}^2/2 + \Phi(r)$. $J_r$ is evaluated using the true spherical
potential out to $r_{\mathrm{term}} = 3 r_{200}$, beyond which (for
reasons of numerical speed) the calculation is truncated and an
analytic completion assuming a Keplerian (vacuum) potential is
taken. We verified that changing $r_{\mathrm{term}}$ to $4 r_{200}$
had little impact on the results.

Before proceeding to a comparison, we need to derive appropriate
$\beta$ values. We calculate these using a Monte-Carlo Markov chain
(MCMC) to maximize the likelihood
\begin{equation}
\mathcal{L}(\vec{\beta}, \beta_E) = \prod_i f(\vec{J}_i; \vec{\beta}, \beta_E)
\end{equation}
where $f$ is the 1-particle distribution function
\eqref{eq:maxent-solution} normalized such that $\int \dd^3 \vec{J}\,
\dd^3 \vec{\Theta} \,f(\vec{J}) = 1$. This normalization must be
accomplished numerically on a grid of $J_r, j$ values; at each
grid-point $E(J_r, j)$ is calculated by operating a bisection search on
equation \eqref{eq:radial-action}. This need only
be done once, and then the evaluation of each link in the Markov chain
is rapid.

The operation gives us maximum likelihood (i.e. ``best fit'')
parameters\footnote{The MCMC technique also yields uncertainties on
  the $\beta$ values, but these will not be considered further in the
  present work.} $(\beta_j, \beta_z, \beta_r, \beta_E)$ for a given
simulation, optionally subject to constraints (such as $\beta_E=0$ or
$\beta_j = \beta_z=\beta_r=0$).  We are now fitting up to four
parameters (excluding mass normalization), more than the one or two
parameters normally used by simulators to describe their halos
\citep[e.g.][]{2004MNRAS.349.1039N,2009MNRAS.398L..21S}. However the
fitted real-space density profiles are purely phenomenological
constructs; conversely here we are starting with a functional form
derivable from physical considerations.
As we have commented in Section \ref{sec:new-canon-ensemble} and will
expand upon in Section \ref{sec:other-simulations}, the $\beta$'s
should ultimately therefore be derived from initial conditions. For
the present, however, the objective is to see whether our physical
argument can correctly describe the phase space distribution at all,
for which fitting $\beta$'s is the most pragmatic approach.

\subsection{Comparison with MW: $J_r$ distribution}\label{sec:results}

We will now start to test how closely equation \eqref{eq:maxent-solution}
represents the distribution of particles in our simulations. We will
investigate MW in some detail, before showing results for the
other two simulations to which the same discussion can essentially be
applied.

The distribution of $J_r$ values in MW is shown by the histogram in
the left panel of Figure \ref{fig:Jr}. This can be compared with the
thick solid curve which shows the distribution of $J_r$ values
according to expression \eqref{eq:MAXENT-Jr-derivative}; the
parameters are $\beta_E^{-1} = 1.6 \times 10^4 \,\ks$ and
$\beta_r^{-1} = 4.4 \times 10^3 \,\kks$.  The agreement is excellent
over several orders of magnitude in probability density, spanning the
values $0<J_r<J_{\mathrm{break}}$ where $J_{\mathrm{break}} \simeq 3
\times 10^4\, \kks$. Particles with $J_r>J_{\mathrm{break}}$ account for
less than $0.1\%$ of the mass and are on long period orbits, probably
reflecting new material falling into the potential well. We will not
consider them further.

The shape of the $J_r$ solution can be understood as follows. We have
already remarked that, in the limit $\beta_E \to 0$, one recovers
equation \eqref{eq:Jr-only-distrib}, an exact exponential (i.e. a
straight line on the linear-log axes of Figure \ref{fig:Jr}). For comparison we have plotted
the best fit distribution of this form (with $\beta_r^{-1} = 3.5
\times 10^3\, \kks$) as a dashed line. Since the
period of an orbit increases with its energy (or radial action), the mean frequency $\langle \Omega_r \rangle_{J_r}$
decreases for increasing $J_r$. So, inspecting equation
\eqref{eq:MAXENT-Jr-derivative}, there will always be a $J_r$ value above
which $\beta_E \langle \Omega_r \rangle_{J_r}$ becomes much smaller
than $\beta_r$. Looking again at the thick solid curve in Figure
\ref{fig:Jr}, the limiting solution at high $J_r$ is indeed a pure
exponential as this reasoning would suggest. At small $J_r$, however,
the gradient of the solution is steeper because of the energy term.

Comparing the histogram, the thick solid curve and the dashed line in
the left panel of Figure \ref{fig:Jr} thus leads us to the conclusion
that $J_r$ conservation (dashed line) accounts rather well for the
qualitative form of the distribution, with an important correction
from $E$ conservation at low $J_r$. Finally the dotted curve shows the
best fit case with $\beta_r=0$ -- i.e. the normal statistical
mechanical result in the absence of other constraints -- and provides
a poor fit at all $J_r$.  In summary, the identification of the $J_r$ constraint has resulted in
dramatic improvements in the match to simulations.

\subsection{Comparison with MW: $j$ distribution}\label{sec:results-j}

\begin{figure}
\includegraphics[width=0.47\textwidth]{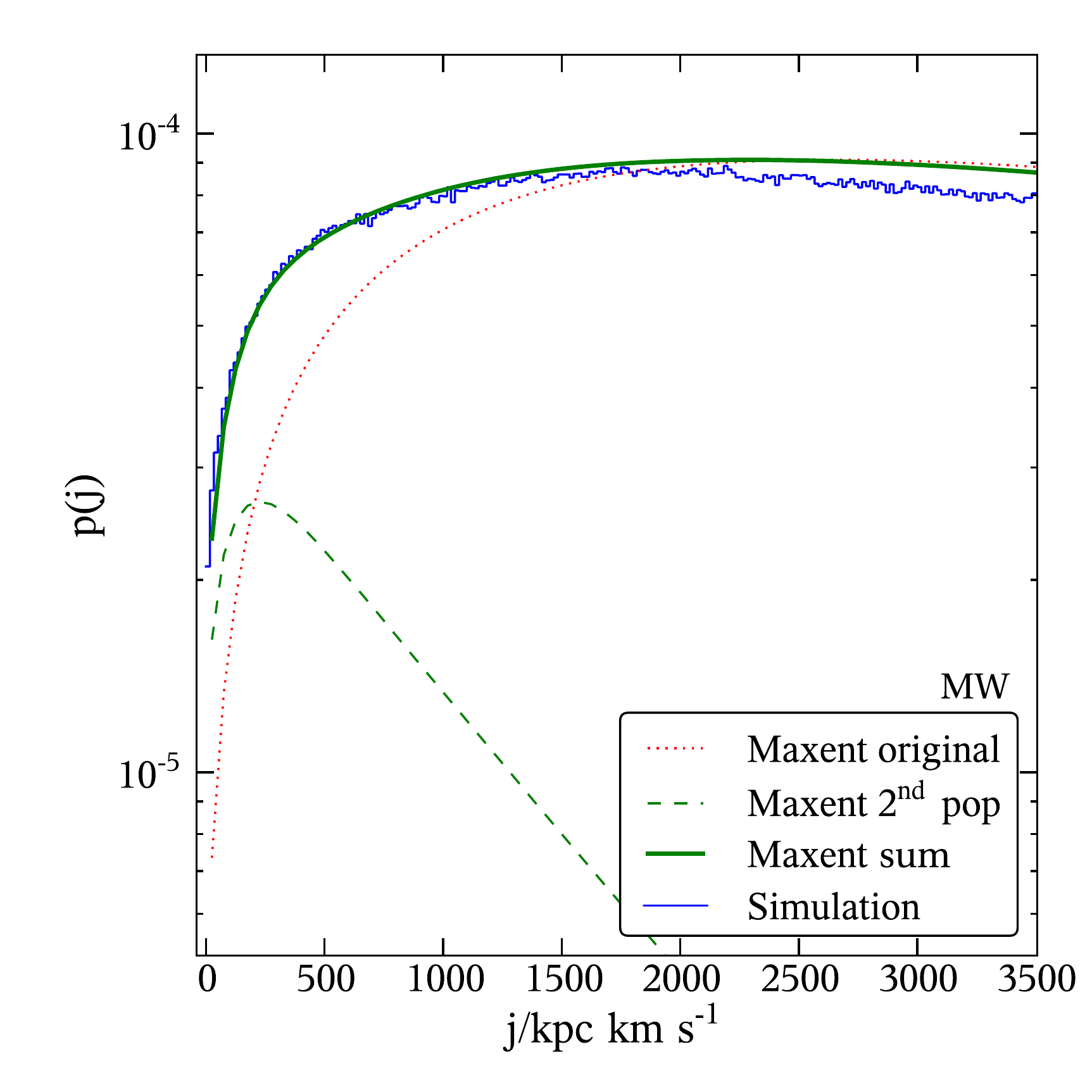}
\caption{Despite good agreement over the majority of $j$ space (see
  right panel of Figure \ref{fig:Jr}), the fraction of simulated
  orbits (histogram) at very low angular momentum is substantially
  underestimated by the simplest maximum entropy argument (thin dotted
  curve). One fix discussed in the text is to postulate a second
  population at low energies (dashed curve). This yields a much better low-$j$ fit
  (solid line) without affecting the high-$j$ fit (except through a
  minor renormalization). }\label{fig:h285-blowup}
\end{figure}

\begin{figure*}
\includegraphics[width=0.9\textwidth]{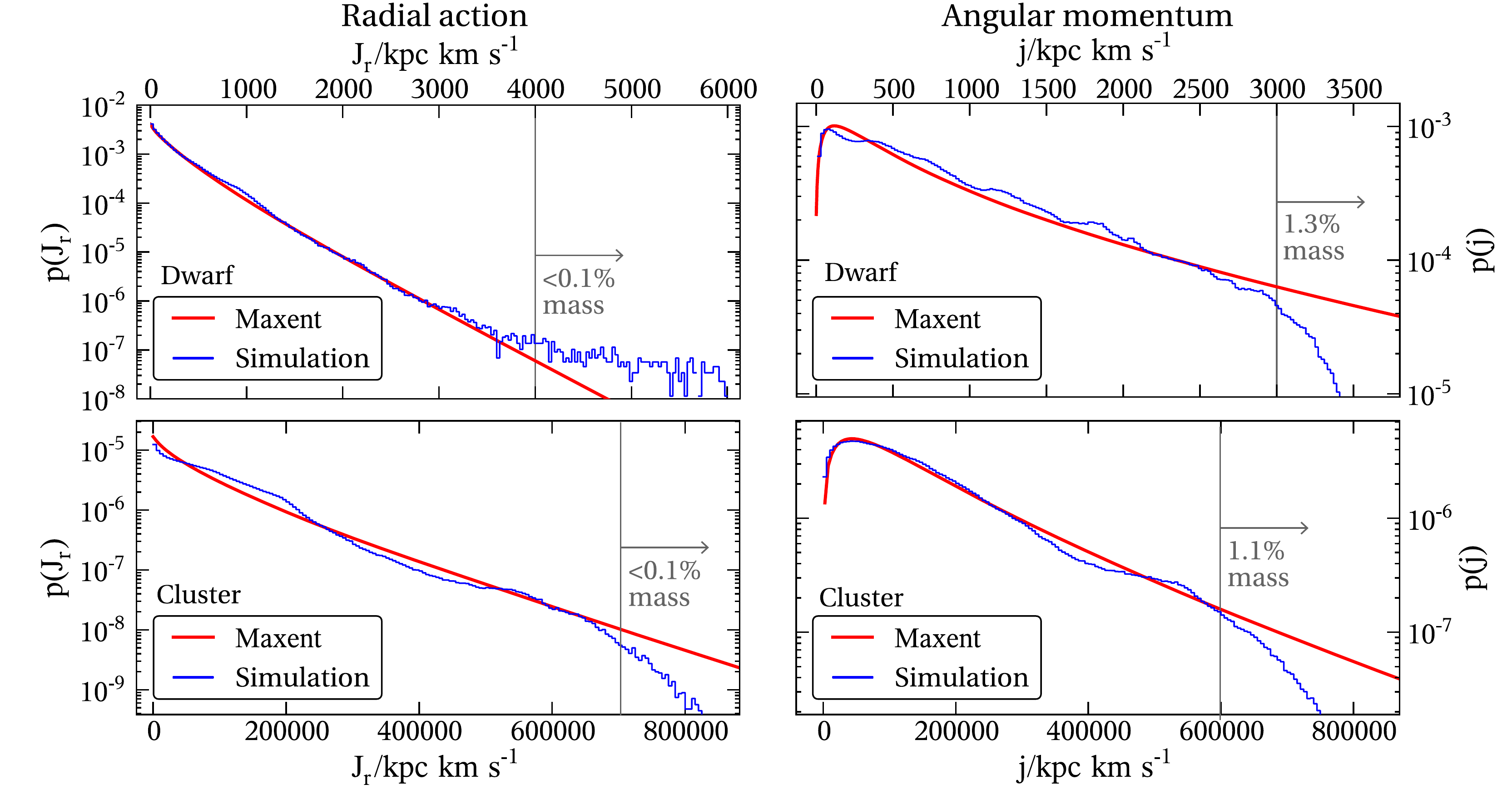}
\caption{As Figure \ref{fig:Jr}, but for the remaining two
  simulations. Once again, the maximum entropy distribution subject to
  $\vec{J}$ and $E$ constraints (thick solid lines) predicts the
  simulations (histogram) accurately. The simulated distribution in
  $J_r$ for the cluster (lower left panel) has some noticable
  fluctuations over large scales; this is likely because it is
  dynamically young.}\label{fig:others}
\end{figure*}

Now consider the right panel of Figure \ref{fig:Jr} which shows the
distribution of scalar angular momentum for the particles in our MW
simulation. Once again the simulated particles are shown by the
histogram; the best fit maximum entropy solution ($\beta_j^{-1},
\beta_{z}^{-1} = 4.4\times 10^3, 1.1\times 10^4\, \kks$, with
$\beta_E$ as quoted above) is shown by the thick solid curve. It again
reproduces the correct qualitative behaviour up to $j_{\mathrm{break}}
= 4 \times 10^4 \,\kks$, with only $0.2\%$ of the mass at
$j>j_{\mathrm{break}}$. Although the angular momentum distribution has
some fluctuations away from the predicted
behaviour, the predictions remain nearly correct over two orders of
magnitude in probability density. With the exception of a problem
described below, we do not believe these fluctuations to be of
particular importance beyond indicating the structure is not
completely relaxed. In particular we will show later (Section
\ref{sec:real-space}) that these inhomogeneities can be ignored when
reconstructing a density profile in real space. Certainly compared
against a solution based on $E$ conservation alone, again shown by a
dotted curve, our solution can be counted a success.

The basic shape of the predicted $j$ distribution can be understood in
a similar way to the $J_r$ distribution explained above.  Consider
again the case where $\beta_E=0$ (so in effect the total energy is
unconstrained); then the exact solution is given by
equation~\eqref{eq:angmom-no-E}. We can also take the isotropic limit,
$\beta_z \to 0$, giving
\begin{equation}
p_j(j) \propto j\, \exp(-\beta_j j)\hspace{1cm}(\beta_z=0,\textrm{ }\beta_E=0).
\end{equation}
This is analogous to the radial action case
\eqref{eq:Jr-only-distrib}, but with a degeneracy factor $j$
reflecting the increasing density of available states available as the
angular momentum vector grows in size. The result is that the abundance
of particles grows linearly with $j$ for $j<\beta_j^{-1}$ and decays
exponentially for $j>\beta_j^{-1}$.

In light of the above discussion, it is notable that the turnover from
growth to decay in $p_j(j)$ occurs at $j$ values much smaller than
$\beta_j^{-1}$. There are two ways to accomplish this. The first is to
create a highly anisotropic setup, $\beta_z^{-1} \ll j_0$, where $j_0$
is the smallest $j$ value of interest. This packs orbits as much as
possible into a single plane, generating a large net angular momentum
and destroying the approximate spherical symmetry\footnote{We note in
  passing that, technically, distribution functions with net angular
  momentum can nonetheless generate spherical potentials
  \citep{1960MNRAS.120..204L}.}, but effectively removing the
degeneracy in $j$ altogether:
\begin{equation}
  p_j(j) \propto \exp \left[(\beta_z-\beta_j) j\right] \hspace{1cm}(\beta_z^{-1}\ll j, \textrm{ }\beta_E=0)\textrm{.}
\end{equation}
Because it is maximally anisotropic, this solution cannot reflect the
simulations; however if we temporarily fit only $j$ values using the functional
form \eqref{eq:angmom-no-E}, we are pushed towards this unphysical
limit (dashed line, Figure \ref{fig:Jr}, right panel;
$(\beta_j^{-1},\,\beta_z^{-1}) = (5.9,\, 6.3)\times 10^2\,\kks$).

Luckily this is not the only way to overcome the
shrinking phase space at low $j$. Equation
\eqref{eq:MAXENT-angmom} shows that if $\langle \Omega_j\rangle_j$
increases fast enough as $j\to 0$ it can overcome the $\coth(\beta_z j)$ degeneracy
term. We have verified that in the
full solution (thick solid line in Figure \ref{fig:h285-blowup} right
panel), this is the mechanism by which the turnover is pushed to low $j$.

Focussing attention on the low-$j$ part of the distribution does,
however, reveal a deficiency in our predictions.
Figure \ref{fig:h285-blowup} shows the distribution of orbits with
$j<3500\,\kks$. The dotted line shows the same maximum entropy fit
depicted by the solid line in Figure \ref{fig:Jr}. When the horizontal
scale is expanded in this way, it becomes clear that the global fit
undershoots the simulation values significantly at low $j$. This
appears to be a systematic feature of all simulations we have
inspected (the three detailed here, GHALO, and various other lower
resolution simulations which we used for testing purposes). It is
possible to force a better fit by restricting the likelihood analysis
to this region, but the global agreement is then considerably worse.

This suggests that the behaviour at low $j$ is marginally decoupled
from that in the rest of phase space. This could arise from the wide
range of orbital periods: particles with small actions also have
periods much shorter than the rest of the halo. The coupling between
particles will necessarily be weak if their timescales are very
different (since particles on short orbits react adiabatically to
fluctuations on long timescales). This can substantially suppress
redistribution of scalar angular momentum and is consistent with,
although not reliant on, the early formation of a stable central cusp
in simulations
\cite[e.g.][]{1998ApJ...499L...5M,2006MNRAS.368.1931L,2011MNRAS.413.1373W}.

In principle this weakness of coupling between orbits in different
regions could be expressed as a further constraint in the maximum
entropy formalism. Further investigation awaits future work, but for
now we will use this as a motivation to study a two-population
system. We are not suggesting that there really are two sharply
defined populations, but that this should anticipate the 
features of incomplete equilibrium. 

Our maximum likelihood analysis is able to find a dramatically better
fit in this case, placing 3.5\% of the mass in a second population at
substantially lower temperature ($\beta_E^{-1} = 2.8\times 10^3
\,\ks$). The summed distribution is shown by the thick solid line in
Figure \ref{fig:h285-blowup}, with the contribution from the
subdominant population indicated by the dashed line. Because the
second distribution is so peaked near $j=0$, the only difference at high $j$
is a marginal renormalization. We also verified that the $J_r$
distribution is barely affected.

It is undeniably disappointing that our solution does not
automatically accommodate the behaviour at very low $j$, but we expect
that future development of the ideas above can quantitatively account
for the discrepancy. We consider other possible explanations in
Section~\ref{sec:conclusions}. However after focussing so much on one
corner of phase space we should re-emphasize the major conclusion: the
distribution over both $J_r$ and $j$ for 96\% of the particles are
remarkably well described by the maximum entropy expression~\eqref{eq:maxent-solution}.

\subsection{Other simulations}\label{sec:other-simulations}

We confirmed that these basic conclusions persist in other
simulations. The top row of figures in Figure \ref{fig:others} shows
results from the `dwarf' simulation. The $J_r$ distribution (top left
panel) again shows excellent agreement for $J_r<J_{\mathrm{break}}$,
where $J_{\mathrm{break}} = 4\times 10^3 \, \kks$. Only a
tiny fraction of mass ($<0.1\%$) lies beyond this point of
breakdown. 
%(One possible point of interest is that, for $J>J_{\mathrm{break}}$
%the simulation seems to have an over-abundance of particles
%rather than the under-abundance seen in the other simulations. Whether
%this has any particular significance is beyond the scope of our
%current considerations.)
As with MW, the dwarf's angular momentum distribution (top right
panel) has more conspicuous fluctuations, but still roughly adheres to
the maximum entropy solution up to $j_{\mathrm{break}}=3 \times
10^3\,\kks$, with around $1.3\%$ of mass lying beyond this point. We
verified that at very low angular momenta $j<100\,\kks$ there is again
an overabundance of particles in the simulation, although in this case
it accounts for less than $2\%$ of particles compared against the
$3.5\%$ in MW.  The parameters of the dwarf fit are $(\beta_r^{-1},
\beta_j^{-1}, \beta_z^{-1}) = (4.2, 1.9, 2.1) \times 10^2\,\kks$ and
$\beta_E^{-1}= 2.2 \times 10^3 \,\mathrm{km\,s^{-1}}$.

\begin{figure}
\includegraphics[width=0.47\textwidth]{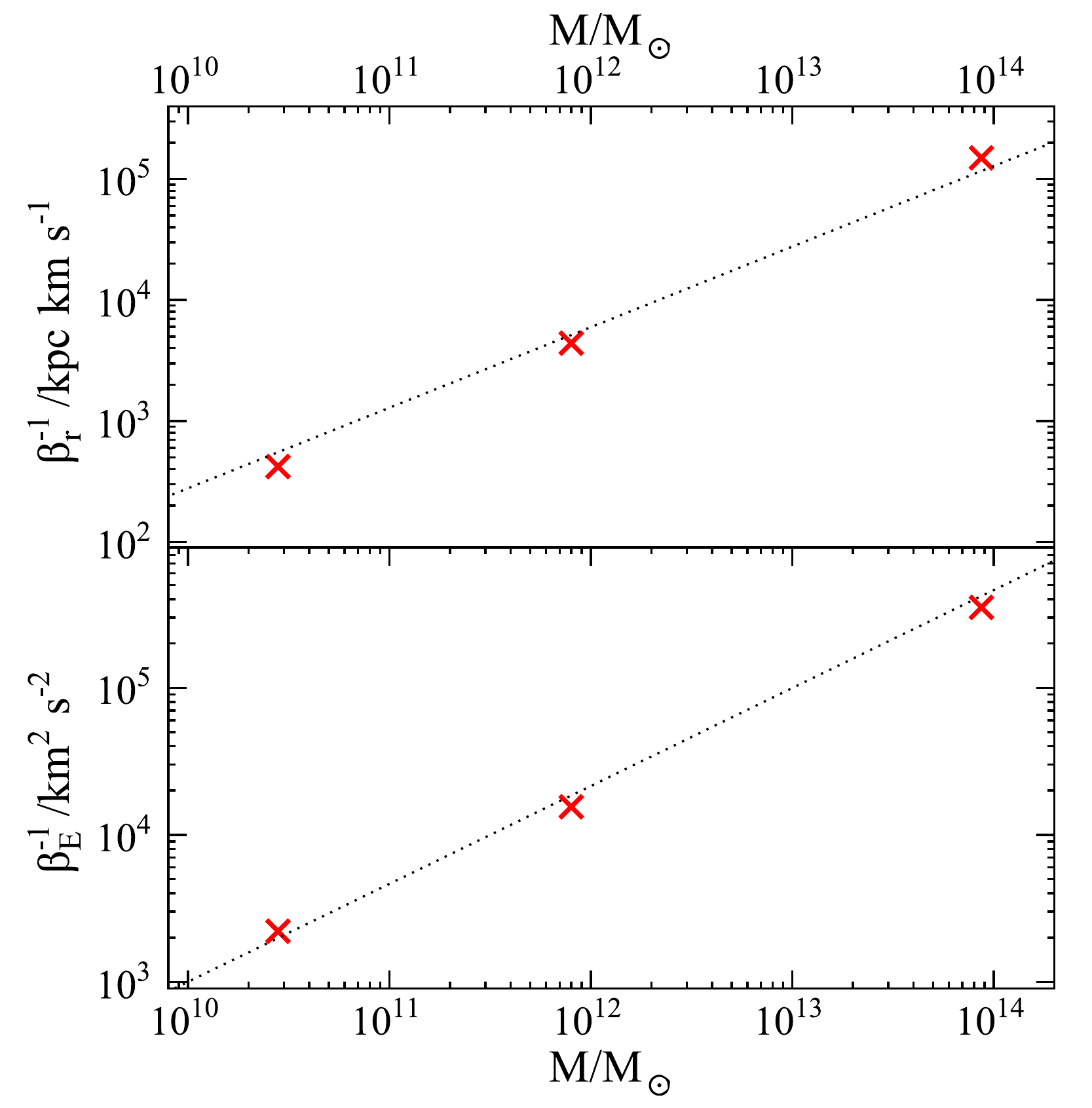}
\caption{The best fit scales for radial action (upper panel) and
  energy (lower panel) as a function of mass. The crosses show the
  values from the three simulations, while dotted lines give the
  expected scalings \eqref{eq:E-scale} and \eqref{eq:J-scale}, which
  agree well with the simulations.}\label{fig:scalings}
\end{figure}

Considering the cluster simulation (lower row of Figure
\ref{fig:others}) gives similar results once again. This time the
$J_r$ distribution as well as the $j$ distribution shows some notable
fluctuations around the maximum entropy description. This may be
because clusters assemble later (as we discussed in Section
\ref{sec:perf-comp-techn}, this is reflected in the lower
concentration value), so the system is dynamically young; however we
have not explicitly looked at time dependence of these distributions.
The parameters of the cluster fit are $(\beta_r^{-1}, \beta_j^{-1},
\beta_z^{-1}) = (1.5, 0.8, 1.2) \times 10^5\,\kks$ and $\beta_E^{-1}=
3.5 \times 10^5 \,\mathrm{km\,s^{-1}}$ with $0.1\%$ and $1.1\%$ of the
mass in the unrelaxed components beyond $J_{\mathrm{break}} = 7\times
10^5\,\kks$ and $j=6\times 10^5\,\kks$ respectively.

Naively one would expect $\beta_E^{-1}$ to scale approximately as
\begin{equation}
\beta_E^{-1} \propto \frac{GM_{200}}{r_{200}} \propto M_{200}^{2/3}\textrm{,}\label{eq:E-scale}
\end{equation}
since $M_{200}$ and $r_{200}$ are by definition related through the
fixed-mean-density condition $M_{200} \propto r_{200}^3$. Similarly the actions
$\vec{\beta}^{-1}$ should scale as
\begin{equation}
\vec{\beta}^{-1} \propto r_{200} \sqrt{\frac{GM_{200}}{r_{200}}} \propto M_{200}^{2/3}\textrm{.}\label{eq:J-scale}
\end{equation}
Figure \ref{fig:scalings} compares these expectations with the actual values,
although we immediately caution against taking the scaling of
three halos too seriously. The upper panel shows the radial
action, $\beta_r^{-1}$, as a function of mass (crosses) with dotted
lines indicating the scaling~\eqref{eq:J-scale}. The lower panel
shows the same for the energy scales. Both panels show good agreement
with the expected trends. 
%There is a hint that the radial action
%(upper panel) scales more steeply than the simple analysis
%given above predicts. This could reflect the varying concentration
%of the halos (Figure \ref{fig:simulations}): as the concentration
%decreases (with increasing mass), more particles are found at larger
%radii, requiring
%characteristic actions to increase faster than equation
%\eqref{eq:J-scale}.

For clarity we did not over-plot the $\beta_j$ values in Figure
\ref{fig:scalings}, but these can be seen to be comparable to
$\beta_r$. Because cosmological halos are formed from near-cold
collapse, their initial angular momentum will be small. The final
dispersion of angular momentum is likely generated through a weak form
of the radial orbit instability
\citep[e.g.][]{1991MNRAS.248..494S,2006ApJ...653...43M,2008ApJ...685..739B,2009ApJ...704..372B}.
Thus the scales of the angular momentum distribution and the radial
action distribution are likely to be intimately linked. This is one
example of a dynamical consideration which should ultimately be used
to link $\vec{\beta}$ values to the initial conditions.

\subsection{Real space radial density profiles}\label{sec:real-space}

\begin{figure}
\includegraphics[width=0.47\textwidth]{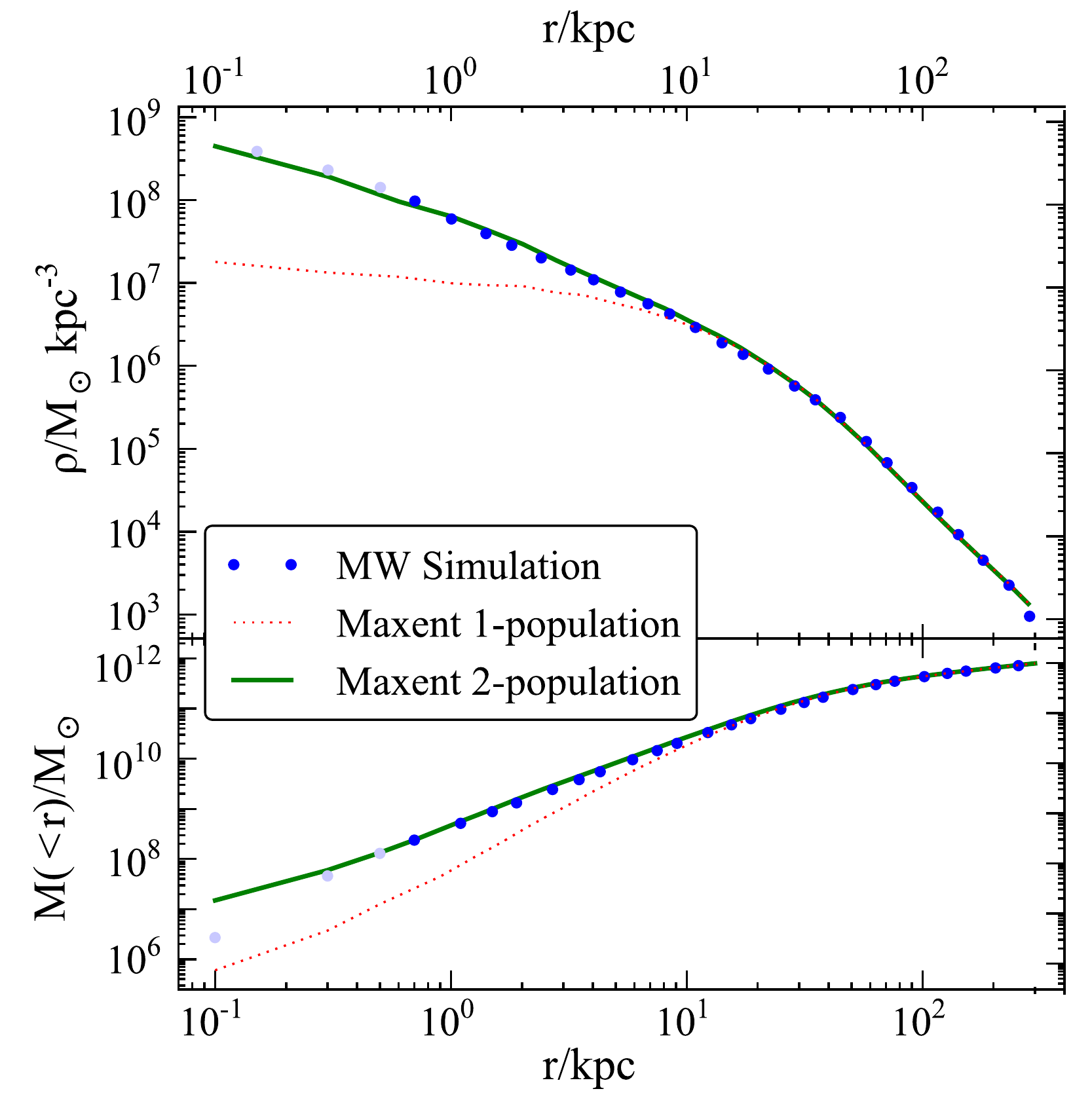}
\caption{The real-space density distribution (upper panel) of the 1-component and
  2-component maximum entropy solutions (dotted and solid lines
  respectively) compared to the MW simulation binned density profiles
  (dots). The softening length in MW is $170\,\pc$, so the profile
  should be reliable exterior to $\sim 700\,\pc$
  \protect\cite[e.g.][]{2003MNRAS.338...14P}. The generic maximum entropy
  result is a density profile with slowly steepening powerlaw to
  increasing radii, in agreement with the simulations. The 1-component
  fit misses the central density cusp, showing that the $\sim 3.5\%$
  correction to the low angular momentum orbits (Section
  \ref{sec:results-j}) is required to reproduce this quintessential
  feature of simulated dark matter halos. The lower panel shows the
  cumulative mass as a function of radius.}\label{fig:rho}
\end{figure}

We have shown that a first-principles maximum entropy argument is
capable of describing the phase space distribution of particles in
dark matter halos, up to a small correction at low angular
momentum. The natural next step is to ask what kind of real-space radial
density profiles are implied by this phase space distribution
and whether these match the classic rolling-powerlaw shape given by
simulations.

Calculating the density distribution corresponding to the phase space
distribution~\eqref{eq:maxent-solution} is technically involved; a description is given in
Appendix \ref{sec:dynam-dens-estim}. There we also explain how the same
computer code can be used to calculate equilibrium density profiles
from simulations (as opposed to analytic distributions). These
profiles are generated subject to our simplifying assumptions of phase
mixing and spherical symmetry. They agree well with traditional
`binned' estimates of the density, validating the
assumptions. Furthermore in the new method, each simulated particle is
smeared out over its orbit, resulting in considerably smaller
Poisson noise than from traditional binned estimates.

Applying the algorithm to the maximum entropy solution, we find that
the radial density profile implied by equation
\eqref{eq:maxent-solution} follows a shallow power law in the centre
and steepens with increasing radius, in qualitative agreement with the
behaviour seen in numerical simulations \citep[][and references
therein]{1997ApJ...490..493N,2004MNRAS.349.1039N,2009MNRAS.398L..21S}.
However, when using the single population phase-space distribution
fits, the central slope is too shallow (dotted line,
Figure~\ref{fig:rho}). One can obtain higher central densities and
inner slopes by changing the parameters, but then the outer slope
becomes too steep. On the other hand, if one adopts the incomplete
relaxation fit advocated in Section \ref{sec:results-j}, a vastly
improved real space density profile is recovered (thick solid line,
Figure \ref{fig:rho}). This confirms that the $\sim 3.5\%$ population
at low-$j$ is responsible for controlling the cusp. In the discussion
below we will recap our current understanding of this issue and give
directions for future investigation.

\section{Discussion}\label{sec:conclusions}

We have shown that maximizing the entropy of a distribution function
subject to constraints on total action and energy reproduces the phase
space density of particles in simulated dark matter halos.  Crucially,
there is a clear physical motivation behind this choice of
constraints. We started by explaining that, since any equilibrium
distribution must be phase-mixed, the late stages of relaxation 
approach this phase-mixed state. As a consequence $\langle \vec{J}
\rangle$ becomes a conserved quantity as equilibrium is approached
(Section \ref{sec:conservation-action}). This constitutes a dynamical
barrier to continued evolution, preventing energy from being further
redistributed. 

The resulting canonical ensemble (i.e. the maximum entropy solution) is given by
equation~\eqref{eq:maxent-solution}.  From it we derived two key
relationships which can be used to test the phase space of simulated
halos, respectively equations~\eqref{eq:MAXENT-Jr-derivative}
and~\eqref{eq:MAXENT-angmom}. These were used to demonstrate a close
agreement between simulations and theory (Section \ref{sec:results},
\ref{sec:results-j}) over orders of magnitude in probability density,
and over a wide range of halo masses from dwarf galaxies to clusters
(Section \ref{sec:other-simulations}). We compared to the
\cite{1967MNRAS.136..101L} distribution which is obtained when energy
can be arbitrarily redistributed between particles, finding that our new canonical ensemble
offers a vastly improved fit (see dotted lines
in Figure~\ref{fig:Jr}). This strongly suggests that ({\it a})
maximum entropy with suitable dynamical constraints (representing
incomplete violent relaxation) is a plausible route to understanding
the 6D phase space of dark matter halos; {\it (b)} the newly
constrained quantities need not be conserved in general, but must be
conserved whenever the system is close to equilibrium, so that their
value becomes fixed as the dynamics settle down; and ({\it c}) we have
identified a physical argument leading to an important example of
these constraints.

However we found an overabundance of low angular momentum
orbits in the simulations relative to the analytic predictions
(Section \ref{sec:results-j}). This implies that there is at least one
more important constraint that we have not fully reflected in our
analysis. Constructing the radial density profile (Section \ref{sec:real-space})
confirms that, although a small fraction ($\lsim 4\%$) of particles
are causing the discrepancy, their existence is essential to
understanding the origin of the central density cusps seen in
numerical simulations.

The correspondingly large density of particles as $j\to 0$ has been
found by previous work, notably \cite{2001ApJ...555..240B}, who
offered a fitting formula which implies continually increasing
particle numbers towards $j=0$. With our higher
resolution simulations, we can see that $p_j(j)$
does eventually decrease at sufficiently low $j$, but slower than
expected given the shrinking available phase space (Figure \ref{fig:h285-blowup}).
Furthermore the large number of particles at low angular momentum can
be linked directly to various discrepancies between $\Lambda$CDM
theory and observed galaxies
\citep{2001MNRAS.326.1205V,2009MNRAS.396..141D}. In particular there
must be mechanism to remove the low angular momentum baryons
\citep{1986ApJ...303...39D,2001MNRAS.321..471B,2010Natur.463..203G,2010arXiv1010.1004B}. Understanding
what causes the accumulation of low angular momentum material in the
first place is now added to the list of puzzles in this area.

Our maximum entropy picture gives an interesting framework in which to
interpret the situation. In Section \ref{sec:results-j} we gave an
extensive analysis of equation \eqref{eq:MAXENT-angmom} which suggests
two routes to adding material at low $j$. The first option is to
appeal to anisotropy (first term on the right hand side); the second
is to use a population of particles at low energy (high $\beta_E$ in
the second term on the right hand side). We currently prefer the
second explanation for the following reason. Particles near the centre
have very short orbital periods, which make them decouple from
fluctuations on the dynamical timescale of the remainder of the halo.
In numerical simulations, the cusps are indeed the first part of the
halo to form, and they do not change much at late times
\citep{1998ApJ...499L...5M,1998MNRAS.293..337S,2011MNRAS.413.1373W}.
\cite{2006MNRAS.368.1931L} construct an explicit 2-phase model of the
formation of halos reflecting this differentiation, emphasizing the
lack of equilibriation between the inner and outer parts of the halo
\cite[see also][]{2011ApJ...743..127L}. Accordingly a timescale
constraint could be incorporated from the outset of the maximum
entropy argument; we expect this would give similar results to our
current approach of fitting a second population.  This will be tackled
explicitly in future work.

%A closely related interpretation is that the negative heat capacity of
%self-gravitating systems leads to the formation of a two-phase system
%with a dense central cusp \citep{1968MNRAS.138..495L}.  Run-away
%transfer between the components is sometimes assumed inevitable in
%such systems, but in the spirit of the current work we would expect
%dynamical constraints to prevent this on any finite timescale. This
%may therefore be another way of looking at the effective 2-phase
%solution we have used in the present work.

The alternative view is that the behaviour at low $j$ may be sensitive
to effects of asphericity. This could modify the effective degeneracy.
But as we commented in Section \ref{sec:results-j}, the only obvious
method available is to pack orbits tightly into a plane, so making the
phase space available uniform with $j$, rather than linearly
increasing. Numerical results do show halos become more anisotropic
towards their centre \citep{2002ApJ...574..538J}. On the other hand,
when given a second population to fit, our code does not select this
as a viable explanation for the existence of the cusp (Section
\ref{sec:results-j}).

If a full description of the physics generating low angular momentum
orbits can be reached, the work in this paper lays the foundation for
a complete description of the collisionless equilibria of dark matter
halos. Further questions of interest will include:
\begin{itemize}
\itemsep 0.2em
\item whether and how the
constraint vector $\vec{\beta}$ can be derived from initial conditions
\citep[which will likely lean heavily on an understanding of the
radial orbit instability,
e.g.][]{1991MNRAS.248..494S,2006ApJ...653...43M,2008ApJ...685..739B,2009ApJ...704..372B};

\item whether and how maximum entropy arguments can explain the power-law
  behaviour of the pseudo-phase-space density $\rho(r)/\sigma^3(r)$
  \citep{2001ApJ...563..483T};

\item whether and how the phase mixing is maintained to sufficient
  accuracy to make the $\langle \vec{J} \rangle$ conservation
  effective over periods of major disturbance \citep[perhaps through
  chaotic mixing e.g.][]{1996ApJ...471...82M,1997PhRvL..78.3426H};

\item how various moments of the distribution function evolve at higher
order in perturbation theory (which is closely related to the
previous question); 

\item how the arguments are changed by adopting an explicitly
  aspherical background Hamiltonian (assuming this is
  not already answered when attacking the low-$j$ question); and

\item how maximum entropy arguments can be
adjoined to microphysical descriptions of cusp destruction
\citep{1996MNRAS.283L..72N,2001ApJ...560..636E,2002ApJ...580..627W,2005MNRAS.356..107R,2006Natur.442..539M,2012MNRAS.421.3464P}
to shed further light on this essential area of galaxy formation.
\end{itemize}

Our substantial step forward should give confidence that a full
statistical account of the distribution of particles in simulated dark
matter halos is achievable without any ad hoc assumptions or
modifications to the well-established principle of maximum entropy. Such an
account would be extremely powerful for practical and pedagogical
aspects of understanding the behaviour of dark matter in the Universe.

\section*{Acknowledgements}

AP gratefully acknowledges helpful conversations with Steven Gratton,
James Binney, Justin Read, Simon White, Carlos Frenk, Julien
Devriendt, James Wadsley, Phil Marshall, Julianne Dalcanton, Jorge Pe{\~n}arrubia and John
Magorrian, and thanks Kieran Finn for development of computer code for
a related project.  The MW simulation was run by Alyson Brooks.  The
GHALO simulation was kindly made available by Joachim Stadel and Doug
Potter. FG was funded by NSF grant AST-0908499.  NSF grant AST-0607819
and NASA ATP NNX08AG84G.  Simulations were run on NASA Advanced
Supercomputing facilities.  Simulation analysis was performed with the
pynbody package ({\tt http://code.google.com/p/pynbody}) on the DiRAC
facility jointly funded by STFC, the Large Facilities Capital Fund of
BIS and the University of Oxford. This work was supported by the
Oxford Martin School and the Beecroft Institute of Particle
Astrophysics and Cosmology.

\bibliographystyle{mn2e} 
\bibliography{../refs.bib}

\appendix

\vfill
\section{Why maximize entropy?}\label{sec:but-what-about}

In this Appendix we return to the question of why we have derived
particle distribution functions by maximizing the
entropy~\eqref{eq:entropy}. We will give an outline of Jaynes'
reasoning \cite[e.g.][]{2003prth.book.....J}: that maximizing entropy
subject to given constraints is equivalent to testing whether those
constraints encapsulate the physics of the situation.

We start by outlining two schools of thought explaining why maximizing
entropy is meaningful. The first relies on the ``H-theorem'' which
states that the entropy increases with time, and hence systems evolve
towards a state which maximizes their entropy. However the entropy
$S[f]$ as defined\footnote{\cite{1967MNRAS.136..101L} discusses an exclusion
  principle which arises from Liouville's theorem and can modify the
  classical expression for entropy; the deviations will be significant
  if the initial phase space density is comparable to the density in
  any regions of a final `coarse-grained' view of phase space. This
  does not seem likely to apply in cosmological settings
  \citep{1978ApJ...225...83S}, although see \cite{2012ApJ...748..144B}
  for a different view.} by equation \eqref{eq:entropy} is actually exactly
constant in time for a collisionless system. A related quantity which
does increase with time is the entropy of the `coarse-grained'
distribution function $F$. Here, $F$ is discretized and equal to $f$
averaged over a local volume in phase space. But there are an infinity
of functionals of $F$ which increase with time, and no obvious reason
to favour $S[F]$ over these alternatives \citep{1986MNRAS.219..285T}.

Similar difficulties extend to collisional systems. In these cases it is only
the Gibbs entropy that is perfectly conserved, and Boltzmann's entropy
typically increases with time \citep{1965AmJPh..33..391J}. However
there are experimentally accessible cases where the Boltzmann entropy
systematically decreases with time
\citep{1971PhRvA...4..747J}. Consequently the justification for
expecting systems to adopt maximum entropy states is not clear from
the H-theorem even in this case.

The second school of thought states that entropy represents human
uncertainty about the state a particle will be found in
\citep{1957PhRv..106..620J}. In this case, the entropy
functional~\eqref{eq:entropy} is derived from Shannon's
axioms\footnote{For a unique answer we also need to demand that
  entropy be invariant
  under coordinate transformations of the phase space, which reflects
  our intial ignorance of the distribution of particles before the
  dynamics are specified. For an alternative view, see \cite{2010ApJ...725..282W}
  who recently suggested that the measure should be uniform in energy
  space. This is equivalent to imposing an
  {\it a priori} preference for some regions of phase space over
  others, one that should ultimately be derived from the equations of
  motion (and could then be re-expressed as a constraint). } -- these are reasonable requirements for what `human
uncertainty' can mean \citep{2003prth.book.....J}. The last of
Shannon's four axioms (which requires additivity of entropy of
independent systems) has been questioned
\citep[e.g.][]{1988JSP....52..479T,Plastino99}. However, no
significant improvement in matching the phase space of simulations has
resulted from these developments \citep{2008PhRvE..77b2106F}.
Moreover, the axiom in question can be viewed as requiring ``no
unwarranted correlations'' \cite[see the section on kangaroos
in][]{Jaynes86Monkeys}, in the sense that specifying constraints on
expectations of any variables $\langle a \rangle$ and $\langle b
\rangle$ will by default choose $\langle a b \rangle = 0$ unless any
other information specifies to the contrary. This seems a strongly
desirable property. The remainder of the discussion therefore focuses
on the known properties of the Boltzmann entropy.

There is one outstanding question: why should we maximize our
uncertainty? It turns out that if we have to choose a state based on
the given constraints, the vast majority of all possibilities (putting
a uniform prior probability on $f(\omega)$ at each point in phase
space) are arbitrarily close to the maximum entropy result
\citep{JaynesEntropyConcentration}. This might be reflected in an
`ergodic' hypothesis that the system actually explores all of these
states, but it is not necessary that this be the case. The key insight
of Jaynes is that if the system is consistently found in a
different state from that predicted, this is evidence for a systematic
effect. Once a physical model of that effect has been built, it can be
incorporated as a further constraint and the maximum entropy formalism
still stands \citep{Jaynes79}.

Hence if we ask ``does maximum entropy subject to these constraints
reproduce the numerical distribution function'' we are really asking
``do these constraints encapsulate the important physics
of dark matter halo collapse?''. That is the aim of this work.

\section{Dynamical Density Estimates}\label{sec:dynam-dens-estim}

\begin{figure}
\includegraphics[width=0.49\textwidth]{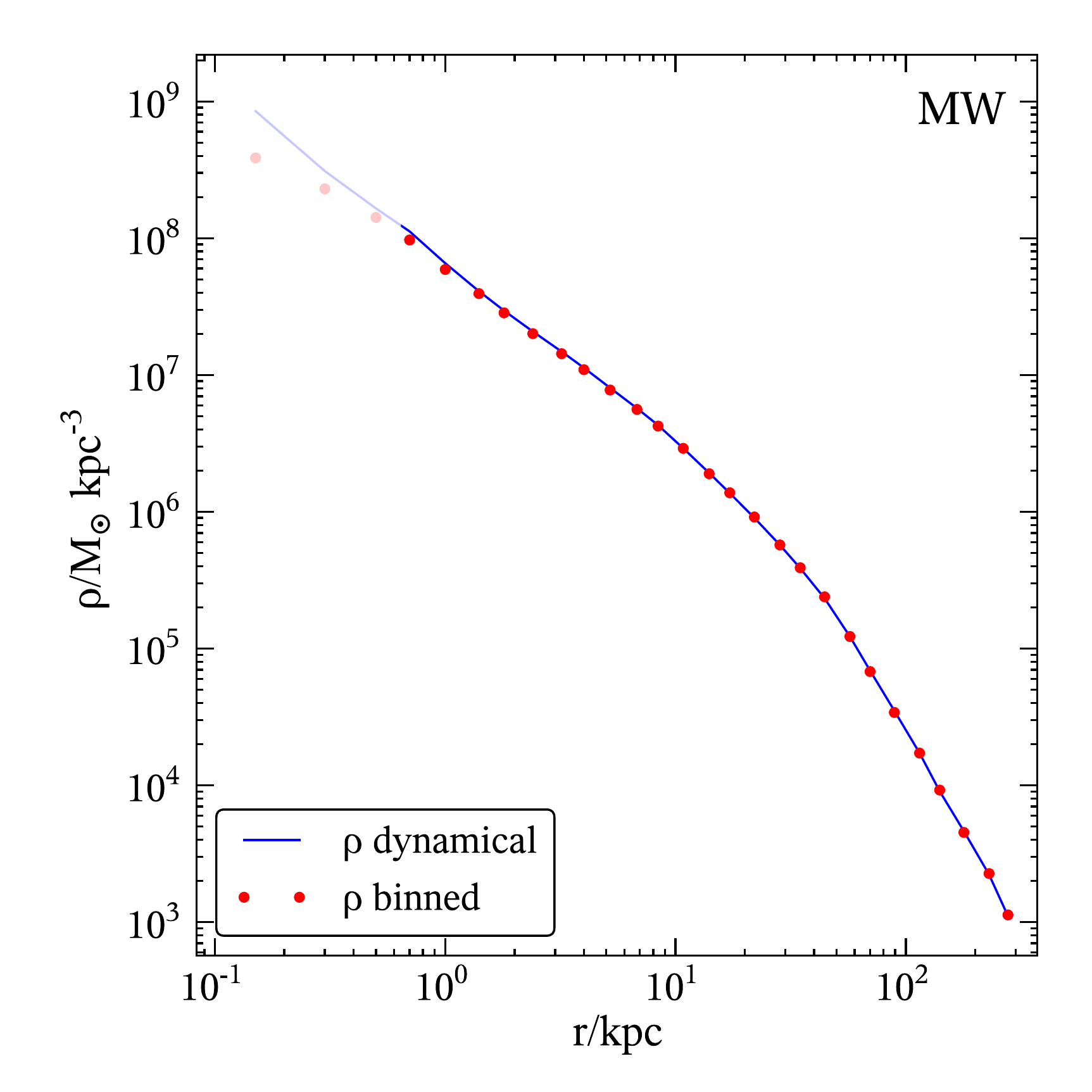}
\caption{As a test of our assumptions about phase mixing, we can
  generate a density profile of the simulated halo MW (solid line)
  where phase information is thrown away and the mass from each particle is consistently `smeared' along
  its orbit. The result is plotted as a solid curve, and can be compared to
  the direct `binned' density estimate of the same simulation; the
  agreement is excellent where the profile is reliable (exterior to
  $4\epsilon \simeq 700\,\pc$).}\label{fig:sanity-check}
\end{figure}

In Section \ref{sec:real-space} we discussed the real space radial
density profiles resulting from our ensemble. We now explain how these
are calculated. Starting from expression~\eqref{eq:maxent-solution}, the mass enclosed
inside a radius $r$ is given by
\begin{equation}
M(<r) = M_0 \iint \dd J_r\, \dd j\,  p(J_r,j)\, P(<r; J_r,
j)\textrm{,} \label{eq:M-equation}
\end{equation}
where $p(J_r, j)$ is the distribution function marginalized over
$j_z$,
\begin{equation}
p(J_r,j) = \int \dd j_z\, f(\vec{J}) = \sinh \beta_z j\,
\exp\left(-\beta_r J_r -\beta_j j -\beta_E E(J_r, j)\right)\textrm{,}\label{eq:p-Jr-j}
\end{equation}
and $P(<r;J_r, j)$ gives the fraction of time that a particle with orbital
parameters $(J_r, j)$ spends interior to radius $r$:
\begin{equation}
P(<r; J,j) \propto \mathrm{Re} \int_0^r \dd r \left(E(J_r,j) - \frac{j^2}{2 r^2} - \Phi(r)\right)^{-1/2}\textrm{.}\label{eq:P-interior}
\end{equation}
Taking the real part circumvents the need to find apocentre or
pericentre explicitly; however the actual numerical evaluation of this
integral presents some difficulties discussed in Appendix
\ref{sec:numerical-problems}.

The final solution $M(<r)$ depends on $\Phi$ [explicitly through
equation \eqref{eq:P-interior} and implicitly through $E(J_r,j)$ in
equation \eqref{eq:maxent-solution}]. A full solution thus demands an
iterative approach. However, we have found that such iteration
presents difficult numerical convergence problems in cases with
$\beta_E\ne 0$, with solutions often oscillating wildly. While we are
working towards a solution to this problem, for the present
investigation it will be enough to use $\Phi(r)$ derived from the
simulation, and ask whether a maximum entropy population would
correctly trace the original density profile. If the answer is `yes'
to reasonable accuracy, the answer will automatically be
self-consistent.

This raises an interesting test case: one can reinsert the actual
simulated $p(J_r,j)$ distribution into the procedure and check that
the results agree with the original density profile. This does not
rely on any of the maximum entropy arguments, but rather tests
numerical algorithms and the assumption that the distribution can be
approximated as spherical and in equilibrium (i.e. phased
mixed). Failure in any aspect would produce density profiles
disagreeing with those obtained from naive binning in real space. 

To test this we take the calculated $(J_r, j)$ values for all
particles in a simulated halo and use these as tracers of the $p(J_r,
j)$ distribution. This throws away all the phase information from the
original simulation. We then construct the dynamical mass
distribution~\eqref{eq:M-equation} using the same method as for the
maximum entropy $p(J_r,j)$.

In Figure \ref{fig:sanity-check} we show the results of this test
applied to the MW halo. The recovered profile (solid curve) is in
excellent agreement with that derived from the raw simulation data
(shown by points) outside the convergence radius $4\epsilon \simeq
700\,\pc$.  This suggests that our analytic assumptions are valid and
the numerical apparatus is working correctly. We have also noted that
in low resolution simulations (not shown) the recovered profile is
significantly smoother than a binned profile. This is because the new
approach averages the profile over a dynamical time; each particle is
smeared through multiple density bins. This could be a useful
technique for mitigating Poisson noise when working with limited
particle numbers.

\section{Corrections at apocentre and pericentre}\label{sec:numerical-problems}

\begin{figure}
\includegraphics[width=0.49\textwidth]{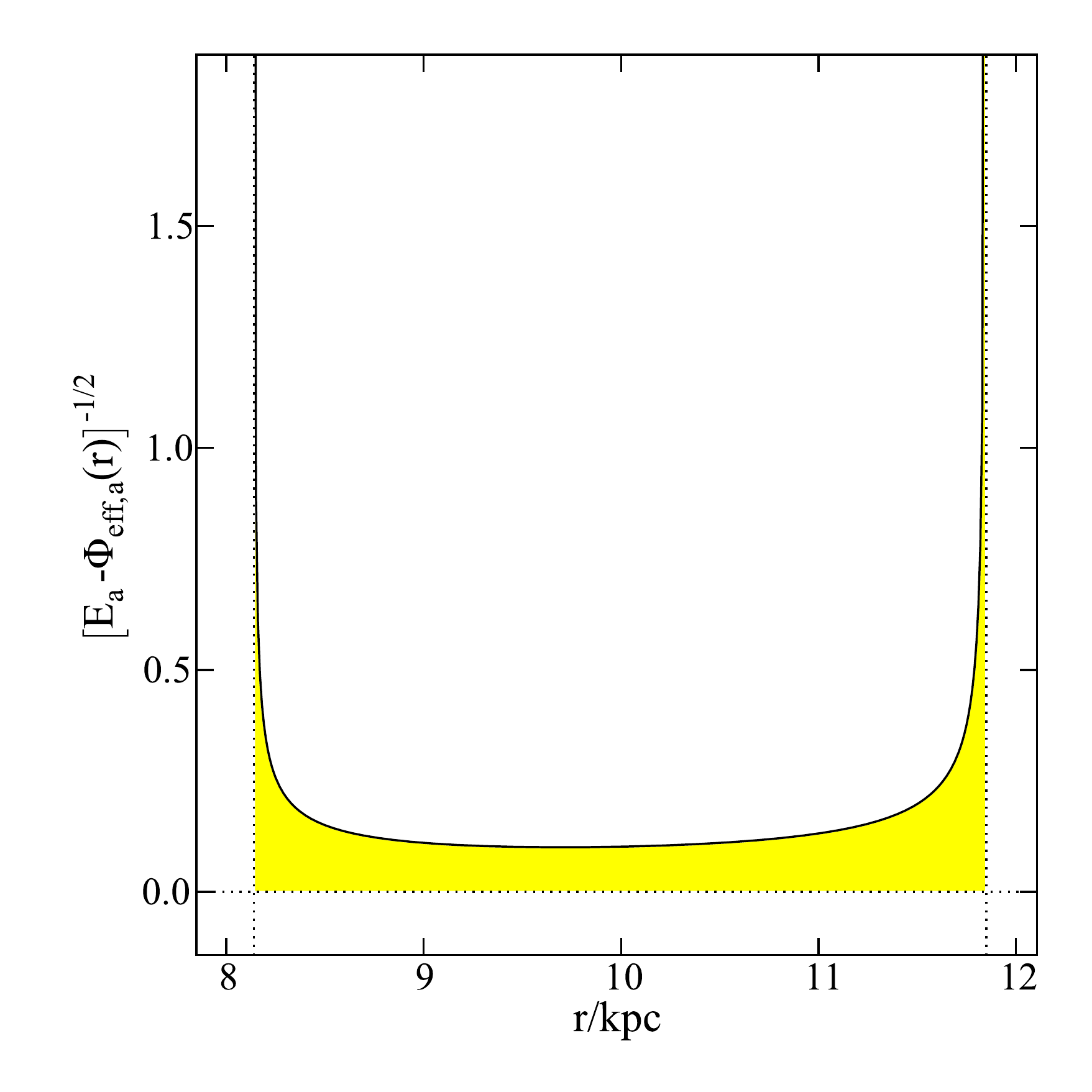}
\caption{A plot of the integrand in equation~\eqref{eq:P-interior} for
  a sample potential, energy and angular momentum. The integrand is
  relatively flat over most of $r$ and can be safely integrated with a
  low-order scheme such as the trapezoid rule. However at apocentre
  and pericentre (here, $\simeq 8.14$ and $11.85\,\kpc$ respectively)
  the integrand diverges. The integral must be evaluated through more
  careful means as explained in the text. }\label{fig:integrand}
\end{figure}
To produce density profiles in real space, as explained in
Appendix~\ref{sec:dynam-dens-estim}, requires rapid, accurate numerical
evaluation of equation \eqref{eq:P-interior}. The
integrand of that expression is plotted for a typical particle in
Figure \ref{fig:integrand}. It is relatively flat over a large range,
and a fast trapezoid quadrature algorithm can therefore be applied. However a branch point at either end of the interval
means that this technique cannot be applied in the endmost
bins. Instead, we use an analytic approximation as described below,
keeping only the lowest order terms in $r-r_0$ where $r_0$ is the branch
point (corresponding to apocentre or pericentre). Consider a particle of energy $E_a$ and angular momentum $j_a$,
and write the effective potential $\Phi_{\mathrm{eff},a}(r) = \Phi(r) +
j_a^2/2r^2$. Then
\begin{equation}
  \left(E_a-\Phi_{\mathrm{eff,a}}(r)\right)^{-1/2}  \simeq 
  \left(-\left.  \frac{\dd
    \Phi_{\mathrm{eff,a}}}{\dd r} \right|_{r_0} (r-r_0) \right)^{-1/2}\textrm{,}
\end{equation}
and so
\begin{equation}
\int_{r_0}^{R} \left(E_a-\Phi_{\mathrm{eff,a}}(r)\right)^{-1/2} \dd r
\simeq  2 \left(\frac{(R-r_0)}{\left.\dd \Phi_{\mathrm{eff},a}/\dd r \right|_{r_0}} \right)^{1/2}\textrm{.}
\end{equation}
We can remove the need to find $r_0$ explicitly by using the relations
\begin{equation}
E-\Phie(R)+(R-r_0) \left. \frac{\dd \Phie}{\dd r}\right|_{R} \simeq 0 \textrm{,}
\end{equation}
\begin{equation}
\left. \frac{\dd \Phie}{\dd r}\right|_{r_0} \simeq \left. \frac{\dd \Phie}{\dd r}\right|_{R}
\end{equation}
to write our final integral approximation as
\begin{equation}
\int_{r_0}^{R} \left(E_a-\Phi_{\mathrm{eff,a}}(r)\right)^{-1/2} \dd r
\simeq  \frac{2 \left(E_a-\Phie(R)\right)^{1/2}}{\left.\dd \Phie/\dd r
  \right|_{R}} \equiv I(R) \textrm{,}
\end{equation}
in which $r_0$ does not appear explicitly. This expression is fast to
evaluate since all quantities are known exactly; the denominator is
just the force $-GM(<R)/R^2 + j/R^3$. 

The integration of equation \eqref{eq:P-interior} over the full range
is accomplished in bins. For a bin $r_0 \to r_1$, we have
\begin{equation}
P(r_0<r<r_1) \propto \mathrm{Re}\left(I(r_1)-I(r_0)\right)\textrm{,}
\end{equation}
even if $r_0$ or $r_1$ lie outside the physical range.
We should apply this approximation in those bins for which it is
more accurate than the trapezium rule. By comparing the leading
errors from both methods, we established the rule that the alternative
integration method described above is used when
\begin{equation}
E_a-\Phie < \left| \frac{\Phie'^4 \Delta r^2}{8\Phie''} \right|^{1/3}\textrm{.}\label{eq:switch-criteria}
\end{equation} 
% !!see note of 24/04 for derivation
where $\Delta r$ is the bin size used for trapezium quadrature. In the
example above, $\Delta r = 10\,\mathrm{pc}$ and
criterion~\eqref{eq:switch-criteria} is satisfied when apocentre or
pericentre is nearer than
$\sim 50\,\mathrm{pc}$ away. Note however
that the errors in the trapezium method, despite being so localized,
become extremely large. Integrating our test case without the
correction leads to $\sim 40\%$ errors in the outermost
$100\,\mathrm{pc}$ density bins
(centred on $8.15\,\kpc$ and $11.85\,\kpc$),
so the effort over this correction is worthwhile.

\end{document}